# Phase Difference based Doppler Disambiguation Method for TDM-MIMO FMCW Radars

Qingshan Shen and Qingbo Wang

College of Computer Science and Technology,
Nanjing University of Aeronautics and Astronautics, Nanjing, China

## ABSTRACT

*State-of-the-art automotive radar sensors use a Mutliple-Input Mutiple-Output (MIMO) approach to obtain a better angular resolution. Time-Division Multiplexing (TDM) scheme is commonly applied to realize the orthogonality in time at the transmitter. Apart from its simplicity in implementation, TDM scheme has the drawback of a reduced maximum unambiguous Doppler proportional to the number of transmitters. In this paper, a phase difference based Doppler disambiguation method is proposed to regain the maximum unambiguous Doppler which is equivalent to only one transmitter. This method works well when the number of transmitters is large. The proposed method is demonstrated with simulation and measurement data.*

## 1. INTRODUCTION

Current-generation automotive radar requires high resolution in the aspect of range, velocity and angle of azimuth. For high-resolution estimation of arrival angle, a wide aperture is indispensable. Frequency-Modulated Continuous Wave (FMCW) radars are commonly used in cars for Advanced Driver Assistant Systems (ADAS) with Mutliple-Input Mutiple-Output (MIMO) technology to meet these requirements. MIMO radar systems consist of multiple transmitters and multiple receivers, offering a large number of virtual antenna elements and relatively high angular resolution [1], [2]. The signals emanating from multiple transmitters need to be orthogonal and this is mainly implemented with the following approaches: Time-Division Multiplexing (TDM), Frequency-Division Multiplexing (FDM), Code-Division Multiplexing (CDM) and Doppler-Division Multiplexing (DDM). TDM-MIMO is the most intuitive and simple way as each transmitter transmits its own waveform alternatively. Ideal orthogonality can be obtained because there is no overlap between any two transmissions [3], [4].

However, TDM-MIMO scheme results in a reduction in the maximum unambiguous velocity that can be measured by the radar. In order to measure velocity, an FMCW radar transmits multiple chirps separated by time interval $T_c$. Phase difference induced by this interval can be used to estimate the velocity of targets. TDM-MIMO FMCW radars with $N_{TX}$ antennae enlarge this time interval to $N_{TX}T_c$. The maximum unambiguous radial velocity can be given as

$$V_{max} = \pm \frac{\lambda}{4N_{TX}T_c}, \quad (1)$$

where $\lambda$ is the wavelength of the radar. Apparently, the maximum unambiguous velocity is reduced by $N_{TX}$ which is a significant drawback if the number of antennae is large.

To regain the true velocity, several disambiguation techniques have been used previously. The Chinese Remainder Theorem (CRT) can be applied for the disambiguation on the foundation of several subsequent measurements with different time intervals [5], [6].However, when the number of transmitters increases, the CRT algorithm requires more complex time interval configurations and has very limited velocity disambiguation capability. The DBSCAN clustering algorithm can be applied to velocity disambiguation in medium PRF radar and achieve more robustness [7].Previously neglected high-order phase terms in the received FMCW radar echo were utilized for extension of maximum unambiguous velocity in [8] and a space-time adaptive processing approach was used for Doppler ambiguity in [9]. These methods are somewhat difficult to implement in practice. Recently, Hypothetical Phase Compensation (HPC) technique has been used frequently. This method compensates the velocity induced by phase shift and then selects the correct hypothesis by comparing the peaks of angle FFT results [10], [11]. However, as the number of transmitters increases, the complexity of the calculation increases and the accuracy decreases.

This paper utilizes the phase difference to solve velocity ambiguity by making full use of the phase change information of multiple transmitters and receivers in TDM-MIMO radars. This method can obtain the same maximum unambiguous velocity as with the use of a single transmitter. In this process, no extra hardware costs will be needed and the requirements of calculation will be very simple. The proposed method is validated with simulations and measurement data collected with a 77GHz FMCW cascaded radar. The remainder of the paper is arranged as follows. In Section 2, we analyse the phase difference used for Doppler disambiguation. In Section 3, we analyse the case of Doppler ambiguity and derive a formula for disambiguation based on phase difference. We validate the method through simulation and measurement data in Section 4 and we conclude in Section 5.

## 2. PHASE DIFFERENCE ANALYSIS

MIMO radar consists of multiple transmitters (TX) and multiple receivers (RX), forming a large virtual array. We can obtain the virtual array signal by performing a two-dimensional FFT (2D-FFT) processing on each TX-RX pair. The range-FFT resolves objects in range and produces a series of bins. A Doppler-FFT is then performed for each range-bin across chirps and thus a signal at a specific range-Doppler bin indicates an object at that range and velocity [5].

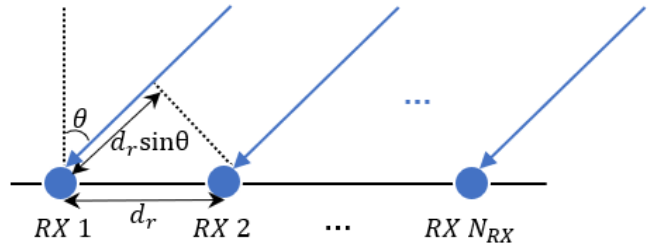

Figure 1. RX antenna array.

Consider an object moving away from radar with a relative velocity $v$, and denote the angle of arrival by $\theta$. Figure 1 shows the RX antenna array. Two adjacent RX antennae are separated by a distance $d_r$, the signal from the object must travel an additional distance of $d_r sin\theta$ to reach the next RX antenna. This distance corresponds to a phase difference between the signals received at two adjacent RX antennae, and the phase difference can be given as

$$\phi_r = \frac{2\pi d_r sin\theta}{\lambda}. \qquad (2)$$

As for TX antennae, the same phase difference can be derived as

$$\phi_{t\_azi} = \frac{2\pi d_t sin\theta}{\lambda}, \qquad (3)$$

where $d_t$ is the distance between two adjacent TX antennae. Since TX antennae transmit alternately in a TDM-MIMO radar, an additional phase difference caused by the velocity of object will be introduced. As the time interval between two adjacent TX antennae is $T_c$, this part of phase difference can be derived as

$$\phi_{t\_v} = \frac{4\pi v T_c}{\lambda}. \qquad (4)$$

The actual phase difference between two adjacent TX antennae can be expressed as

$$\phi_t = \phi_{t\_azi} + \phi_{t\_v} = \frac{2\pi d_t sin\theta}{\lambda} + \phi_{t\_v}. \qquad (5)$$

Combine (2) and (5), the effect of the angle of arrival can be eliminated and thus we can obtain

$$\phi_{t\_v} = \phi_t - \frac{d_t}{d_r}\phi_r. \qquad (6)$$

## 3. DOPPLER DISAMBIGUATION

The relative velocity of the object can be estimated from the Doppler-FFT and denote this velocity by $v_{det}$. This value can be converted into a phase change

$$\phi_{det} = \frac{4\pi v_{det} N_{TX} T_c}{\lambda}, \qquad (7)$$

which corresponds to the phase difference of the same TX antenna between two adjacent chirps. Figure 2 illustrates this phase difference for a positive velocity and a negative velocity.

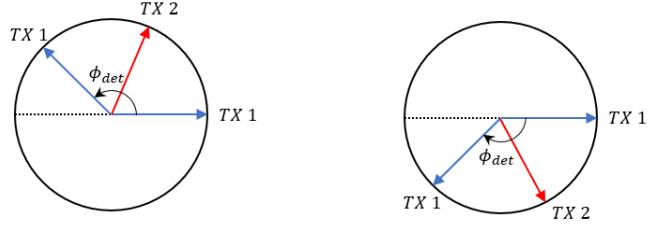

Figure 2. Phase change with a positive velocity (left) and a negative velocity (right).

Doppler ambiguity occurs when the actual phase change exceeds $\pm\pi$. Figure 3 illustrates the actual phase change in the case of two TX antennae and this value can be derived as

$$\phi_{true} = \phi_{det} + 2n\pi, n \in [1 - N_{TX}, N_{TX} - 1] \qquad (8)$$

for $N_{TX}$ TX antennae. And this true phase difference can also be obtained as

$$\phi_{true} = N_{TX}\phi_{t\_v} . \qquad (9)$$

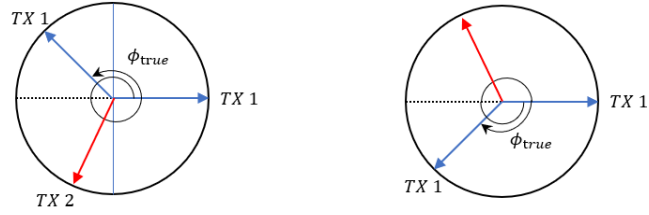

Figure 3. Actual phase change for the situation of two TX antennae, positive velocity (left) and negative velocity (right).

With the combination of (8) and (9), the number of rotations of the phase change can be calculated as follows

$$n = \frac{1}{2\pi}\left[N_{TX}\left(\phi_t - \frac{d_t}{d_r}\phi_r\right) - \phi_{det}\right]. \qquad (10)$$

Once we get the value of $n$, the true velocity can be obtained as

$$v = v_{det} + 2\,round(n)v_{max}. \qquad (11)$$

## 4. RESULTS

### 4.1. Simulation

The data used to validate the method are based on synthetic data generated using the Matlab Radar Toolbox. Parameters of waveform and the corresponding detection capabilities used in this part are shown in Table. 1.

Computer Science & Information Technology (CS & IT) 49

Table 1. Parameters of waveform

| Center Frequency (GHz) | 77 |
|---|---|
| Wavelength (mm) | 3.9 |
| Bandwidth (MHz) | 750 |
| Chirp Time (us) | 42.67 |
| Number of ADC samples | 256 |
| Number of chirps per TX | 64 |
| Maximum Velocity (m/s) | 1.9 |

We employ an antenna array of 12 TX antennae and 8 RX antennae in order to verify the availability of this method with a large number of TX antennae. Figure 4 shows the velocity results enlarged by this method. The object directly in front of the radar moves at the velocity varying from -24 m/s to 24 m/s in step of 0.2 m/s. As can be seen from Figure 5, the value of $n$ is very close to the desired result because a large amount of data is provided for estimating.

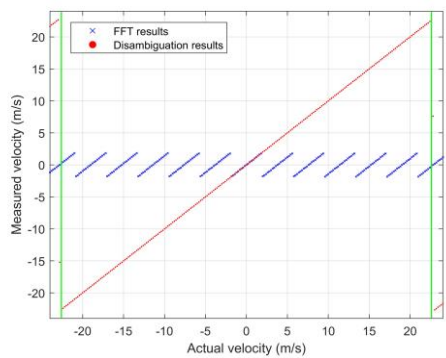

Figure 4. Velocity disambiguation results for an object moving from -24 m/s to 24 m/s, the green lines represent the extended maximum measurable velocity.

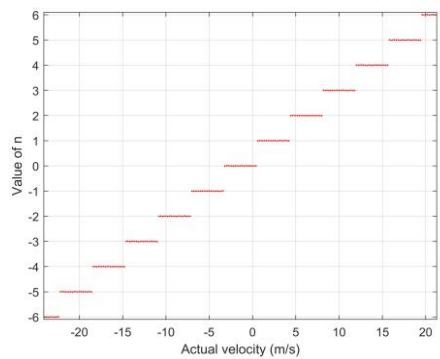

Figure 5. Estimated results of $n$ in simulation.

For situations where the target is not directly in front of the radar, this method can also be available. Figure 6 illustrates the phase information for an object with a velocity of 10 m/s and an angle of 5 degrees to the radar. The RX antennae record information on the azimuth of the object, while the TX antennae record both azimuth and velocity information. Consider an object moving directly ahead at a velocity of 10 m/s, maintaining a certain angle $\theta$ to the radar. The radial velocity of the object can be obtained by multiplying by $cos\theta$. Sweep the angle from -80 degrees to 80 degrees. The unambiguous velocity results are shown in Figure 7 and the velocity is always retrieved correctly.

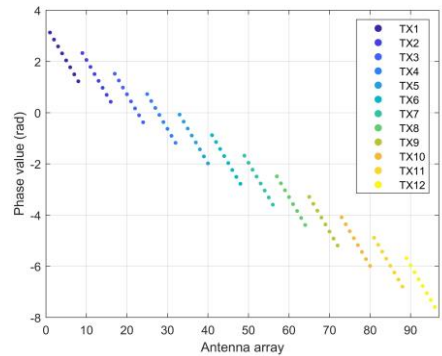

Figure 6. Phase of an object with velocity of 10 m/s and angle of 5 degrees.

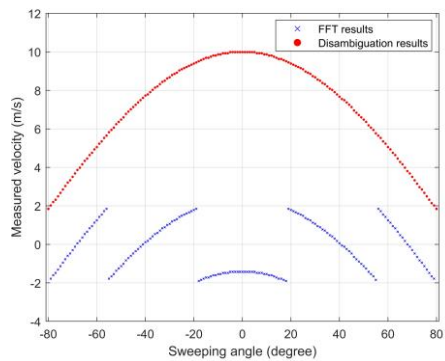

Figure 7. Velocity disambiguation results for sweeping angle.

## 4.2. Measurement Data

The real data is collected with a commercial 77 GHz cascaded radar from Texas Instruments. This module implements a four-device cascaded array of AWR2243 devices and enables support for up to 12 TX and 16 RX antenna elements. It is worth noting that 3 TX antennae of the cascaded radar are used to measure pitch angle. We can only perform velocity disambiguation by using 9 TX antennae in the horizontal direction. Although the maximum velocity is reduced by a factor of 12, it can still be fully recovered with this method.

The environment for data collection is shown in Figure 8, with many houses and trees on either side of the road, which may cause interference. We collected 50 frames of data with the cascaded radar operating in TDM-MIMO mode. During these frames, the car in front of the radar accelerates from 0 m/s to 8 m/s. As can be seen from Figure 9, all velocity values have been successfully resolved. Figure 10 indicates that the value of $n$ always falls around the correct integer and fluctuations for the reason of noise and clutter. From measured data it can be concluded that this method performs well in real data.

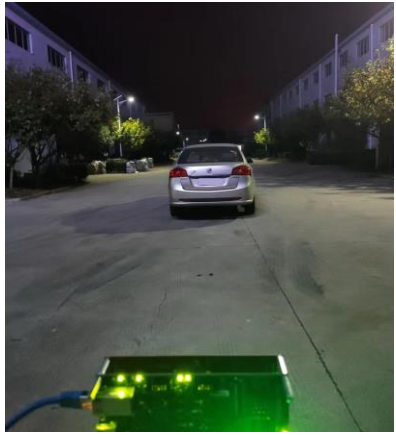

Figure 8. Environment for data collection.

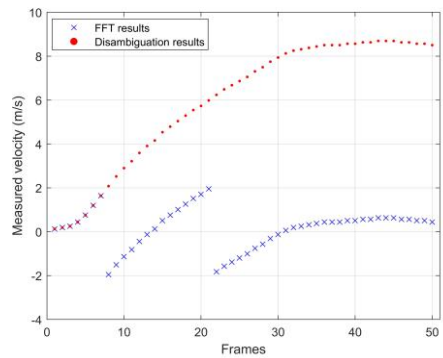

Figure 9. Velocity disambiguation results for a car accelerates from 0 m/s to 8 m/s.

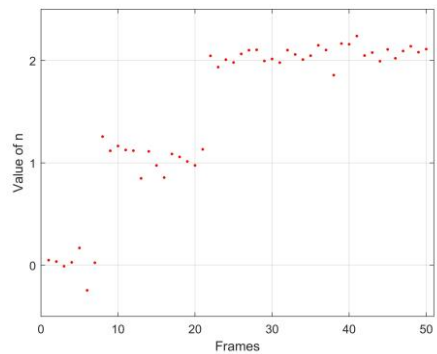

Figure 10. Estimated results of $n$ in measurement data.

## 5. CONCLUSIONS

This paper proposes a Doppler disambiguation method based on phase difference for TDM-MIMO FMCW radars. This method relies on the measurement of the phase and requires a considerable reliability of the phase estimate. The phase difference provided by RX antennae can be used for compensating the angle of arrival in TX antennae. As a result, we can correctly obtain the unambiguous velocity from different TX antennae. This method makes full use of the data from multiple antennae and does not require any additional conditions such as changing chirp

52	Computer Science & Information Technology (CS & IT)times or building overlapping elements in the virtual aperture. This approach is particularly effective when the number of antennae is large. Simulation with Matlab Radar Toolbox and measurement results with cascaded radar are presented to validate this method.

**ACKNOWLEDGEMENTS**

This work was funded by Special Innovation Project for National Defense 19-163-11-ZT-002-002-02. The authors would like to thank the anonymous reviewers for their insightful comments and suggestions.**REFERENCES**

[1]   Sun, S., Petropulu, A. P., & Poor, H. V. (2020). MIMO radar for advanced driver-assistance systems and autonomous driving: Advantages and challenges. *IEEE Signal Processing Magazine, 37*(4), 98-117.
[2]   Roos, F., Bechter, J., Knill, C., Schweizer, B., & Waldschmidt, C. (2019). Radar sensors for autonomous driving: Modulation schemes and interference mitigation. *IEEE Microwave Magazine, 20*(9), 58-72.
[3]   Sun, H., Brigui, F., & Lesturgie, M. (2014, October). Analysis and comparison of MIMO radar waveforms. In *2014 International Radar Conference* (pp. 1-6). IEEE.
[4]   Rao, S., Subburaj, K., Wang, D., & Ahmad, A. (2020). *U.S. Patent No. 10,627,483*. Washington, DC: U.S. Patent and Trademark Office.
[5]   Zhen-xing, H., & Zheng, W. (1987, April). Range ambiguity resolution in multiple PRF pulse Doppler radars. In *ICASSP'87. IEEE International Conference on Acoustics, Speech, and Signal Processing* (Vol. 12, pp. 1786-1789). IEEE.
[6]   Kronauge, M., Schroeder, C., & Rohling, H. (2010, June). Radar target detection and Doppler ambiguity resolution. In *11-th International Radar Symposium* (pp. 1-4). IEEE.
[7]   Tuinstra, T. R. (2016, July). Range and velocity disambiguation in medium PRF radar with the DBSCAN clustering algorithm. In *2016 IEEE National Aerospace and Electronics Conference (NAECON) and Ohio Innovation Summit (OIS)* (pp. 396-400). IEEE.
[8]   Dikshtein, M., Longman, O., Villeval, S., & Bilik, I. (2021). Automotive Radar Maximum Unambiguous Velocity Extension via High-Order Phase Components. *IEEE Transactions on Aerospace and Electronic* Systems.
[9]   Wang, G., & Mishra, K. V. (2020, September). Stap in automotive mimo radar with transmitter scheduling. In *2020 IEEE Radar Conference (RadarConf20)* (pp. 1-6). IEEE.
[10]  GRoos, F., Bechter, J., Appenrodt, N., Dickmann, J., & Waldschmidt, C. (2018, April). Enhancement of Doppler unambiguity for chirp-sequence modulated TDM-MIMO radars. In *2018 IEEE MTT-S International Conference on Microwaves for Intelligent Mobility (ICMIM)* (pp. 1-4). IEEE.
[11]  Liu, C., Gonzalez, H. A., Vogginger, B., & Mayr, C. G. (2021, January). Phase-based Doppler Disambiguation in TDM and BPM MIMO FMCW Radars. In *2021 IEEE Radio and Wireless Symposium (RWS)* (pp. 87-90). IEEE.

## AUTHORS

**Qingshan Shen** was born in Anhui, China, in 1997. He received the B.S. degree in engineering mechanics from Nanjing University of Aeronautics and Astronautics, Nanjing, China, in 2019. He is currently pursuing the Master's degree in Nanjing University of Aeronautics and Astronautics, Nanjing, China. Currently, the major research focus is on signal processing for high resolution millimetre wave radar.

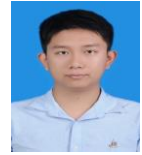

**Qingbo Wang** was born in Jiangxi, China, in 1997. He is currently pursuing the Master's degree in Nanjing University of Aeronautics and Astronautics, Nanjing, China. His main research interests are in the estimation of acceleration parameters for millimetre wave radar.

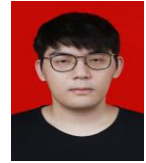